\documentclass[traditabstract]{aa}
\usepackage{txfonts}
\usepackage{graphicx}
\usepackage{natbib}
\usepackage[utf8]{inputenc}
\usepackage{epstopdf}
\usepackage{url}

\newcommand{\mh}{$h^{-1}$Mpc }
\newcommand{\mhp}{$h^{-1}$Mpc}
\newcommand{\vmh}{h^{-1}\mathrm{Mpc} }
\newcommand{\vgh}{h^{-1}\mathrm{Gpc} }

\begin{document}

\title{SDSS DR7 superclusters}
\subtitle{The catalogues}

\author{L.J. Liivamägi\inst{1,2} \and E. Tempel\inst{1,3} \and E.Saar\inst{1,4}}

\institute{Tartu Observatory, Tõravere 61602, Estonia \\ \email{juhan.liivamagi@ut.ee} \and
Institute of Physics, University of Tartu, Tähe 4, Tartu 51010, Estonia \and
National Institute of Chemical Physics and Biophysics, Rävala 10, Tallinn 10143, Estonia \and
Estonian Academy of Sciences, Kohtu 6, Tallinn 10130, Estonia}

\date{Received 09/12/2010, accepted 22/11/2011}

\abstract{We have constructed a set of supercluster catalogues for the galaxies from the SDSS survey main and
luminous red galaxy (LRG) flux-limited samples. To delineate superclusters, we calculated luminosity density
fields using the $B_3$-spline kernel of the radius of 8 \mh for the main sample and 16 \mh for the LRG sample and define
regions with densities over a selected threshold as superclusters, while utilising almost the whole volume of both
samples. We created two types of catalogues, one with an adaptive local threshold and a set of catalogues with different
global thresholds. We describe the supercluster catalogues and their general properties. Using smoothed bootstrap, we
find uncertainty estimates for the density field and use these to attribute confidence levels to the catalogue objects.
We have also created a test catalogue for the galaxies from the Millennium simulation to compare the simulated and
observed superclusters and to clarify the methods we use. We find that the superclusters are well-defined systems, and
the properties of the superclusters of the main and LRG samples are similar. We also show that with adaptive local
thresholds we get a sample of superclusters, the properties of which do not depend on their distance from the observer.
The Millennium galaxy catalogue superclusters are similar to those observed.}


\maketitle

\section{Introduction}
The large-scale structure of the galaxy distribution is characterised by large voids and by a complex web of galaxy
filaments and clusters. Superclusters are the largest components of the cosmic web. They are collections of
galaxies and galaxy clusters, with typical sizes of 20--100 \mhp. They can contain up to hundreds of galaxy groups and
several rich clusters. The first described supercluster is the Local Supercluster \citep{devauc1953}, and many other
superclusters have been found and studied in our neighbourhood.

Astronomers have a long tradition of selecting galaxy clusters and groups from this web, but quantifying the overall web
is a much more difficult task. This can be done in several ways, all of them computer-intensive and based on the
properties of a smoothed galaxy density field. Good recent examples are the application of the multiscale morphology
filter by \citet{jones2010} and the Bayesian inference for the density and the subsequent classification of the web
elements by \citet{jasche2010}. These articles contain an exhaustive set of references. In this approach, the different
sets of web components differ mainly in their dimensionality (clusters, filaments, sheets, and voids). Another approach
that has been used is to divide the observed weblike galaxy distribution into its main building blocks --
superclusters. Superclusters are frequently treated in a similar way to groups and clusters of galaxies -- they are
density enhancements in the overall galaxy distribution.

This approach leads to the construction of supercluster catalogues on the basis of both Abell clusters \citep{maret1997,
maret2001} and galaxy groups \citep{einasto2007}, using smoothed density fields. A similar method has recently been used
by \citet{costa2011} and \citet{luparello2011}. The friends-of-friends method was used by \citet{basilakos2003} to
compile superclusters from the SDSS sample.

Supercluster catalogues are similar to other astronomical catalogues, because while serving as a basis to describe and
study classes of objects, they are also essential for further work. This includes planning observational projects,
comparing different classes of astronomical objects, and comparing theory (simulations) with observations. We present
here the supercluster catalogues based on the richest existing redshift survey, the SDSS DR7. These catalogues have
already been used for several studies. The list includes a study of the locations of quasars within the large-scale
structure delineated by galaxies \citep{lietzen2009}, a couple of observing proposals to search for the warm-hot
intergalactic medium, and a morphological study of the rich superclusters forming the Sloan Great Wall
\citep{maret2010}. This catalogue has also been used for a preliminary identification of a Sunyaev–Zel'dovich (SZ)
source seen in the early Planck mission data \citep{planck2011}.

The paper is organised as follows. In Sect.~\ref{sec:method} we describe our method (beginning with the calculation of
the density field), outline supercluster delineation principles, and explain how some of the more important properties
of the superclusters are calculated. In this section we also address the errors of the density field estimates. In
Sect.~\ref{sec:data} we describe the datasets used. Supercluster properties are described in Sect.~\ref{sec:df_and_scl},
where we also compare different samples. The resulting catalogue can be downloaded from:
\url{http://atmos.physic.ut.ee/~juhan/super/} with a complete description in the \verb1readme1 files. We will also
upload selected parts of the catalogues (listed in Appendix~\ref{app:desc}) to the Strasbourg Astronomical Data Center
(CDS)\footnote{Supercluster tables will be available at the CDS via anonymous ftp to cdsarc.u-strasbg.fr (130.79.128.5)
or via \url{http://cdsweb.u-strasbg.fr/cgi-bin/qcat?J/A+A/}}.

\section{Delineating superclusters by the luminosity density field}
\label{sec:method}

We define superclusters on the basis of their total density that is dominated by dark matter. Supposing that the bias
(the ratio of the dark matter density to the stellar density) is approximately constant on supercluster scales, the
observational counterpart for the total density is the luminosity density. We do not use clusters or groups to create
the density field, as done for earlier supercluster catalogues by \citet{einasto2003,einasto2007}, but the full galaxy
distribution. Before calculating the density field we processed the galaxy data to reduce several observational
selection effects. The galaxy and group samples we used are described in Sect.~\ref{sec:data}.

\subsection{Distance and luminosity corrections for the SDSS main sample}

The spectroscopic galaxy samples (as the SDSS) are affected by the cluster-finger redshift distortions (the
fingers-of-god). To suppress the cluster-finger redshift distortions, we use the rms sizes of galaxy groups and their
radial velocity dispersions from the \citet{tago2010} galaxy group catalogue. In this catalogue, the
comoving distances \citep[see e.g.][]{martinez2003} are used for galaxies and groups, in units of \mhp.
For groups with three or more members, we divide the radial distances between the group galaxies and group centres
($d_{\mathrm{group}}$) by the ratio of the standard deviations $\sigma_r / \sigma_v$. This will remove the smudging of
the density field by the cluster fingers. The corrected galaxy distance $d_\mathrm{gal}$ is found as
\begin{equation}
    d_{\mathrm{gal}} = d_{\mathrm{group}} + (d_{\mathrm{gal}}^\star - d_{\mathrm{group}}) \; \frac{\sigma_r}{\sigma_v /
H_0},
\end{equation}
where $d_\mathrm{gal}^\star$ is the initial distance of the galaxy, $\sigma_r$ the standard deviation of the
projected distance in the sky from the group centre, $\sigma_v$ the standard deviation of the radial velocity
(both in physical coordinates at the group location), and the Hubble constant $H_0 = 100h$ km s$^{-1}$ Mpc$^{-1}$.

We use a cartesian grid based on the SDSS angular coordinates $\eta$ and $\lambda$, because it allows the most efficient
placing of the galaxy sample cone inside a box. The galaxy coordinates are calculated as follows:
\begin{equation}
\begin{array}{l}
    x = -d_{\mathrm{gal}} \sin\lambda, \nonumber\\[3pt]
    y = d_{\mathrm{gal}} \cos\lambda \cos \eta,\\[3pt]
    z = d_{\mathrm{gal}} \cos\lambda \sin \eta.\nonumber
\end{array}
\label{eq:xyz}
\end{equation}

To compensate for selection effects and to ensure that the reconstructed density field does not depend on the distance,
we have to take the luminosities of the galaxies into account that drop out of the survey magnitude window. We follow
the procedure by \citet{tempel2011} and consider every galaxy as a visible member of a density enhancement (a group or
cluster) within the visibility range at the distance of the galaxy. We estimate the amount of unobserved luminosity and
weigh each galaxy as
\begin{equation}
    L_\mathrm{gal,w} = W_L(d)\; L_{\mathrm{gal}},
    \label{eq:gal_weigh}
\end{equation}
where $L_{\mathrm{gal}} = L_\odot 10^{0.4(M_\odot - M)}$ is the observed luminosity of a galaxy with the absolute
magnitude $M$, and $M_\odot$ is the absolute magnitude of the Sun. The quantity $W_L(d)$ is the distance-dependent
weight factor: the ratio of the expected total luminosity to the luminosity within the visibility window:
\begin{equation}
    W_L(d) = \frac{\int_0^\infty L\;\phi(L)\mathrm{d}L}{\int_{L_1(d)}^{L_2(d)} L\;\phi(L) \mathrm{d}L},
    \label{eq:weight_main}
\end{equation}
where $L_{1,2}(d)$ are the luminosity limits corresponding to the survey magnitude limits $M_{1,2}$ at the distance $d$.

We approximate the luminosity function by a double power law:
\begin{equation}
    n(L)\mathrm{d}(L) \propto (L/L^*)^\alpha (1 + (L/L^*)^\gamma)^{(\delta - \alpha)/\gamma} \mathrm{d}(L/L^*),
    \label{eq:lumfun}
\end{equation}
where $\alpha$ is the exponent at low luminosities ($L/L^*) \ll 1$, $\delta$ the exponent at high luminosities ($L/L^*)
\gg 1$, $\gamma$ a parameter that determines the speed of the transition between the two power laws, and $L^*$ the
characteristic luminosity of the transition. This form represents the bright-magnitude end of the luminosity function
better than the usual Schechter function \citep{tempel2009}.

\subsection{Luminosity corrections for the LRG sample}

Although the luminosity function of the SDSS LRGs has already been determined \citep{eisenstein2006}, it is difficult to
calculate the luminosity weights for LRGs as we did above for the main sample. The reason is simple -- the LRG sample
does not have the two magnitude limits. Because of that, we find the observed comoving luminosity density $\ell(d)$
and defined the luminosity weight as its inverse:
\begin{equation}
    W_L(d)=\ell(d_0)/\ell(d),
    \label{eq:weight_lrg}
\end{equation}
where $d_0$ is the fiducial comoving distance (taken as 435.6 \mhp, see Sect.~\ref{sec:lrg}).

Both these luminosity correction schemes (for the main and LRG samples) add luminosity to the observed galaxy
locations, and cannot restore the real, unobserved galaxies. This evidently increases the shot noise at distances, but
that is unavoidable.

\subsection{Calculation of the luminosity density field}

We describe the mathematical details for calculations for the luminosity density field in Appendix~\ref{app:kern}; here
we give a brief summary of the procedure. We denote the luminosity density field on a grid with $\ell_{\mathbf{i}}$,
where $\mathbf{i} = (i_1,i_2,i_3)$ are the indices of the vertices. The luminosity densities are calculated by a kernel
sum:

\begin{equation}
    \ell_{\mathbf{i}} = \frac1{a^3}\sum_\mathrm{gal}  K^{(3)}\left(\frac{\mathbf{r}_\mathrm{gal} -
    \mathbf{r}_\mathbf{i}}{a}\right) L_\mathrm{gal,w},
    \label{eq:dfield}
\end{equation}
where $L_\mathrm{gal,w}$ is the weighted galaxy luminosity, and $a$ the kernel scale.

We use the $B_3$ spline kernel $B_3(x/a)$ (see Appendix~\ref{app:kern}) to construct the one-dimensional kernel
$K^{(1)}(x/a)$, and form the three-dimensional kernel as a direct product of three one-dimensional kernels. The scale
$a$ can be regarded as the effective radius of the kernel, and its choice is determined by the application.

As the last step before extracting superclusters, we convert densities into the units of mean density. The main purpose
of this is to facilitate comparison between different density fields. For that, we construct a pixel mask that follows
the sample edges.
We determine the mean density as an average over all vertices inside the mask,
\begin{equation}
    \ell_{\mathrm{mean}} = \frac1{N_{\mathrm{mask}}}\sum_{\mathbf{i}\in \mathrm{mask}} \ell_{\mathbf{i}},
\end{equation}
where $N_{\mathrm{mask}}$ is the number of grid vertices inside the mask.
We finally normalise the density field as
\begin{equation}
    D_\mathbf{i} = \frac{\ell_\mathbf{i}}{\ell_{\mathrm{mean}}},
\end{equation}
for all vertices with coordinates $\mathbf{i}$ inside the mask. The vertices outside the mask are not used again.

We find the variances $\sigma_{\ell}^2$ of the density field estimates for all vertices by smoothed bootstrap, as
described in Appendix~\ref{app:uncert}. Using that, we calculate for every grid vertex the signal-to-noise ratio
\begin{equation}
    G_\mathbf{i} = \frac{\ell_\mathbf{i}}{{\sigma_{\ell}}_\mathbf{i}}.
    \label{eq:snfield}
\end{equation}
It is used later to estimate confidence levels of superclusters. The parameters and properties of the luminosity
density fields are described in Sect.~\ref{sec:df_prop}.

\subsection{Assembly of superclusters}
We define our superclusters using the luminosity density field. A conventional way is to choose a density level and to
define superclusters as connected density regions above that level \citep[see, e.g.,][]{einasto2007,
luparello2011}. For different tasks, these levels are chosen differently. Because of that, we create sets of contour
surfaces for different density thresholds $D_n$, sampling the density range from $D_{\mathrm{min}}$ to
$D_{\mathrm{max}}$ with a constant increment $\delta D$.

We use density peaks to identify superclusters (density field objects). Contiguous supercluster regions are grown
pixel-wise around the peaks in the density field resulting in a marker field
\begin{equation}
    M_{n,\mathbf{i}} = \mathrm{ID}_\mathrm{peak},\;\mathbf{i} \in \{\mathbf{i} | D_\mathbf{i} > D_n \},
    \label{eq:contfield}
\end{equation}
where $\mathrm{ID}_\mathrm{peak}$ is the density peak number. All the vertices belonging to an object are assigned the
same mark value.

We start scanning the field at high densities and move on to lower density levels. Each time an object first appears, it
is assigned a unique identification number that will be used for this supercluster throughout the catalogue. We keep
track when an object emerges from the field and how or if it is eventually swallowed up by another density field object.
If such a merger occurs, the identifier of the object with the higher peak value will be used to designate that object
later on. To record the merging history of the density field objects, we order them into a tree structure encompassing
all the density thresholds.

We finally assemble superclusters by distributing galaxies among the density field objects. We do this for each density
threshold by correlating galaxy positions with the corresponding marker field. For the SDSS main sample we also assign
galaxy groups to superclusters. If a group or a cluster is found to be in a supercluster (its centre is located inside
the supercluster contour), all its member galaxies automatically also belong to the same supercluster. We also implement
a lower limit of $(a / 2)^3$ for a volume of a supercluster, where $a$ is the smoothing scale, in order to remove small
spurious density field objects that include no galaxies.

\subsection{Selection of density thresholds}
With the multitude of available thresholds comes the question~-- which is the ``correct'' one? Just as there is no
clear-cut definition for superclusters, there is also no single answer for this. We offer two possibilities for tackling
this problem. The first one is the conventional way of choosing a fixed density level, as done above. This gives a set
of objects that are comparable within the whole sample volume, where the density level $D_n$ can be selected according
to the properties of superclusters one wishes to study. As an example, for identifying structures, low density levels
are better, but for studying the details of the structure, higher levels are useful; and sometimes it is necessary
to use a set of luminosity levels. Examples include the density level 5.0 used by \citet{maret2011}, level 4.6 used by
\citet{einasto2007}, 5.5 in \citet{luparello2011}, and the set of levels in \citet{lietzen2009}. However, this approach
is susceptible to Poisson noise, especially in sparser environments. It also does not take the richness differences of
superclusters into account. We demonstrate both effects in Sect.~\ref{sec:scl_prop}.

Because of that, we offer an alternative procedure that assigns an individual threshold to each supercluster, adapting
to the local density level. The idea is to follow the growth of individual superclusters from a compact volume around
its
centre, by lowering the density level and observing the supercluster mergers. By defining a supercluster as the volume
within the density contour until the first merger, we can break the large-scale structure into a collection of compact
components. Every component (supercluster) then has its own limiting density level $D_\mathrm{scl}$, as is usual for
other astronomical objects. We do not define galaxies by a common limiting stellar density level. As a result, we get a
set of superclusters that forms the connected large-scale cosmic web.

To identify such superclusters in practise, it is easier to begin from lower densities and to proceed upwards. The
mergers can now be seen as breakups of structures. We trace the splitting events in the density field objects tree. With
a split a lower density filament ceases to be a ``bridge'' between two higher density regions. We pick the density value
just above of the bridge, after the split, as the defining density level for these two objects. If one of these objects
is broken up again at some higher threshold, it will not affect the other one.

As a downside this technique still requires manually setting several limits. First, the minimal size of a supercluster
must be selected, for obviously some of the breaks involve objects that are too small to be of interest. In previous
studies, a 100 $(\vmh)^3$ lower volume limit was used by \citet{einasto2007}, \citet{costa2011} use ten galaxies as a
minimum for their superclusters (in combination with the volume limit of 64 $(\vmh)^3$). \citet{luparello2011} use the
object luminosity of $10^{12}$ $L_\odot$ as the lower limit. In this study we use the diameter of the supercluster.

We must also choose the maximum threshold $D_\mathrm{lim}$. While we observe that most of the superclusters are defined
at similar density levels, some very rich clusters with their surroundings can satisfy the minimum size condition at a
much higher level and the algorithm may break up well-established structures (we discuss these differences in
Sect.~\ref{sec:scl_prop}). Because of that, we proceed in two steps. First we find the thresholds for all objects, and
we find the maximum threshold $D_\mathrm{lim}$ for superclusters as the density level where 95\% of objects have a lower
threshold. Then we recalculate the thresholds but prohibit splitting of structures above that threshold.

The natural lower density limit is the percolation density level. Percolation happens when the largest structure
starts to fill the sample volume. In practise we define the percolation level $D_{\mathrm{perc}}$ as the density when
the richness of the second richest structure starts to decrease when lowering the density level \citep{martinez2003}.

Shifting the maximum density threshold upwards will fragment structures further. Reducing the minimum size of a
supercluster will have the same effect and will also increase the number of small objects.

We present the SDSS main and LRG supercluster catalogues in two versions, one with a set of fixed levels and the second
with adaptive density thresholds. We describe the differences of these catalogues in more detail in
Sect.~\ref{sec:scl_prop}.

\subsection{Supercluster properties}
After delineating superclusters as described in previous sections, we calculate a number of supercluster properties for
all density levels using both the density field and the galaxy data. In the following we describe the calculation of the
most important attributes of superclusters that will be included in the catalogues. The initial density peak, from which
the supercluster grew and which usually indicates the presence of a large galaxy cluster, marks the supercluster
position.

The supercluster volume is found from the density field as the number of connected grid cells multiplied by the cell
volume:
\begin{equation}
    V_{\mathrm{scl}} = N_{\mathrm{cell \in scl}} \Delta^3,
    \label{eq:vol}
\end{equation}
where $\Delta$ is the grid cell length. We find also the sum of normalised densities at the grid vertices within the
supercluster
\begin{equation}
    L_{\mathrm{scl,df}} = \sum_{\mathbf{i}\in \mathrm{scl}} \ell_\mathbf{i} \cdot \Delta^3,
    \label{eq:ldf}
\end{equation}
for an estimate of the total luminosity of the supercluster. 

Using galaxy luminosities, we obtain two more estimates for the total luminosity of the supercluster, the sum of the
observed galaxy luminosities, and the sum of the weighted galaxy luminosities:
\begin{equation}
    L_{\mathrm{scl,gal}} = \sum_{\mathrm{gal} \in \mathrm{scl}} L_{\mathrm{gal}},
    \label{eq:lum}
\end{equation}
\begin{equation}
    L_{\mathrm{scl,wgal}} = \sum_{\mathrm{gal} \in \mathrm{scl}} L_{\mathrm{gal,w}}.
    \label{eq:wlum}
\end{equation}

We find the number of galaxies, $N_\mathrm{gal}$ and, if available, the number of galaxy groups and
clusters $N_\mathrm{gr}$ in a supercluster. We define the supercluster diameter $\diameter$ as the maximum distance
between its galaxies. Using the galaxy locations and their weighted luminosities, we find the supercluster centre of
mass (luminosity):
\begin{equation}
    \mathbf{r}_{\mathrm{scl}} = \frac{1}{L_{\mathrm{scl,wgal}}} \sum_{\mathrm{gal} \in \mathrm{scl}}
    \mathbf{r}_{\mathrm{gal}}\cdot W_L (d_{\mathrm{gal}})L_{\mathrm{gal}}.
    \label{eq:massc}
\end{equation}

Among these quantities, the most important are the supercluster diameter and the weighted total luminosity, because they
are least affected by the distance to the supercluster. We assume we have restored the real total luminosities by
weighting the galaxies, and while we may lose dim galaxies, the brighter ones still mark the supercluster region
sufficiently \citep{tempel2011phd}. Also, as we show later, neither of these properties is very sensitive to the choice
of the density threshold. Because of the different weighting in the LRG sample, the weighted luminosity there cannot be
used as an approximation to the total luminosity, as for the main sample.

We identify a supercluster by a ``marker galaxy'' that we arbitrarily choose to be a bright galaxy near the highest
density peak in the supercluster volume. The aim of this is to tie a supercluster to an observational object and to
construct an identifier that is not specific to the current catalogue. The long identification number is given in the
format of $AAA \pm BBB+CCCC$, where $AAA$ and $BBB$ are the integer parts of the equatorial coordinates $\alpha$ and
$\delta$ of the marker galaxy and $CCCC$, its redshift multiplied by 1000.

We check whether a supercluster is in contact with the mask edge. A location near the sample boundary implies
incompleteness of the supercluster, and its parameters may not reliable. Using the signal-to-noise field $G$
(Eq.~\ref{eq:snfield}), (see also Appendix~\ref{app:uncert}), we calculate for each supercluster a confidence estimate
\begin{equation}
    C_{\mathrm{scl}} = \frac{1}{N_{\mathrm{gal}}} \sum_{\mathrm{gal} \in \mathrm{scl}} G(\mathbf{r}_{\mathrm{gal}}).
\end{equation}
We interpolate the signal-to-noise ratio values of the density estimate to the galaxy locations and find the average
over all galaxies in the supercluster. An extended description of supercluster properties in the catalogue is given in
Appendix~\ref{app:desc}.

\section{Galaxy and group data}
\label{sec:data}
\begin{figure}
    \input{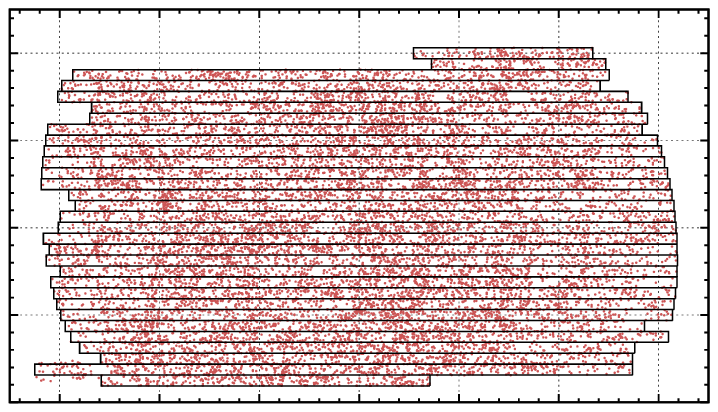}
    \caption{\footnotesize The sky projection of the DR7 galaxies and the mask in the SDSS $\eta$ and $\lambda$ survey
    coordinates.}
    \label{fig:mask}
\end{figure}
We constructed catalogues for both the SDSS main and LRG samples. The main sample has a high spatial density and
allows to follow the superclusters in detail, but the LRG sample, although sparse, is much deeper.

\subsection{The SDSS main sample}
Our main galaxy sample is the main sample from the 7th data release of the Sloan Digital Sky Survey \citep{abaz2008}. We
used the data from the contiguous 7646 square degree area in the North Galactic Cap, the so-called Legacy
Survey (Fig.~\ref{fig:mask}). The sample selection is described in detail in the SDSS DR7 group catalogue paper by
\citet{tago2010}. We used galaxies with the apparent $r$ magnitudes  $12.5 \leq m_r \leq 17.77$ and excluded duplicate
entries. We corrected the redshifts of galaxies for the motion relative to the CMB and computed comoving distances of
galaxies using the standard cosmological parameters: the Hubble parameter $H_0=100 h$ km s$^{-1}$ Mpc$^{-1}$, the matter
density parameter $\Omega_{\rm m} = 0.27$, and the dark energy density parameter $\Omega_{\Lambda} = 0.73$.

\begin{figure}
    \input{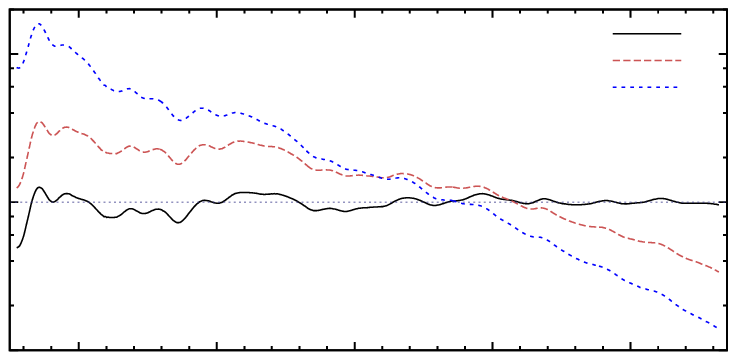}
    \input{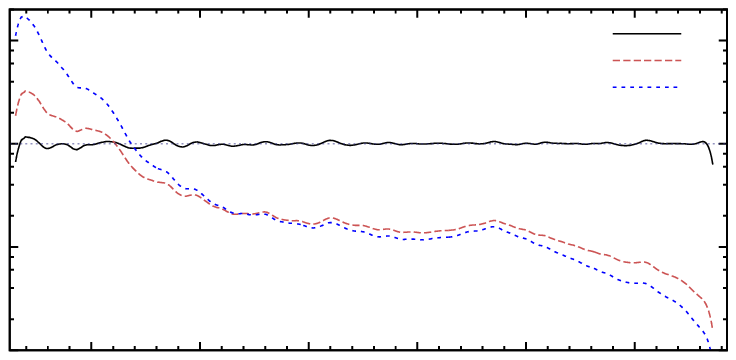}
    \caption{\footnotesize Average normalised densities vs distance for the main ({\it upper panel}) and LRG samples
    ({\it lower panel}). The densities are averaged over thin (a few \mhp) concentric shells of the distance $d$.
    Solid line -- the weighted luminosity density; dashed line -- the observed luminosity density; dotted line -- the
    galaxy number density.}
    \label{fig:densdist}
\end{figure}

We calculated the absolute magnitudes of galaxies in the $r$-band as $M_r = m_r - 25 - \log_{10}d_L - K$, where $m_r$
is the Galactic extinction corrected apparent magnitude, $d_L = d(1+z)$ is the luminosity distance ($d$ is the 
comoving distance) in $h^{-1}\,$Mpc and $z$ the redshift, and $K$ is the $k + e$ correction. The $k$-correction for the
SDSS galaxies was calculated using the KCORRECT algorithm \citep{blanton03a,blanton2007}. In addition, we
corrected the magnitudes for evolution, using the luminosity evolution model of \citet{blanton03b}. The magnitudes
correspond to the restframe (at the redshift $z = 0$).

\begin{figure}
    \input{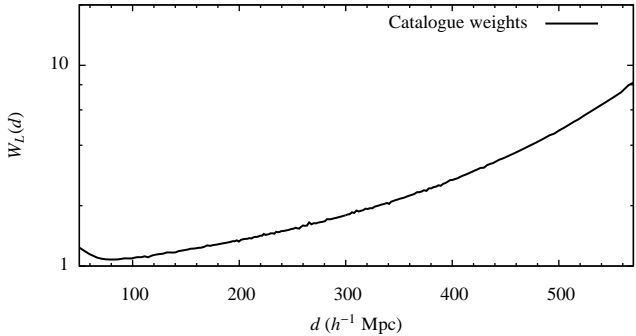}
    \caption{\footnotesize Distance-dependent weights for the main sample galaxies.}
    \label{fig:bootwt}
\end{figure}

Groups and clusters of galaxies were determined using a modified friends-of-friends algorithm, in which a galaxy belongs
to a group of galaxies if this galaxy has at least one group member galaxy closer than the linking length. To
take selection effects into account when constructing a group catalogue from a flux-limited sample, we increased the
linking length with distance, calibrating the scaling by shifting nearby groups \citep[see][for
details]{tago2010}. As a result, the sizes and velocity dispersions of our groups are similar at all
distances. Our SDSS main galaxy sample contains 583362 galaxies and 78800 galaxy groups and clusters.

For the main sample, we use the apparent magnitude limits $m_1 = 12.5$ and $m_2 = 17.7$ for the luminosity limits
$L_{1,2}$ in Eq.~(\ref{eq:weight_main}), and calculate the distance-dependent weight. We take $M_\odot = 4.64$ mag in
the $r$-band \citep{blanton2007} as the luminosity of the Sun. For the luminosity function (Eq.~\ref{eq:lumfun}) we use
the following parameter values: $\alpha = -1.42$ is the exponent at low luminosities ($L/L^*) \ll 1$, $\delta = -8.27$
is the exponent at high luminosities ($L/L^*) \gg 1$, $\gamma = 1.92$ is a parameter that determines the speed of the
transition between the two power laws, and $L^*$ (corresponds to $M^*$ = -21.97) is the characteristic luminosity of the
transition \citep{tempel2011}. Figure~\ref{fig:densdist} shows the dependence of the galaxy number density, the observed
luminosity density, and the weighted luminosity density of galaxies on distance. Our weighting procedure  has adequately
restored total luminosities, the luminosity density does not depend on distance. The luminosity weights are shown in
Fig.~\ref{fig:bootwt}.

\subsection{The SDSS LRG sample}
\label{sec:lrg}
The galaxies for the LRG sample were  selected from the SDSS database by an SQL query requiring that the PrimTarget
field should be either TARGET\_GALAXY\_RED or TARGET\_GALAXY\_RED\_II. We demanded reliable redshifts
($\mathrm{SpecClass} = 2$ and $\mathrm{zConf} > 0.95$). We kept the galaxies within the same mask as the main galaxies
(the compact continuous area in the Northern Galactic Cap). We calculated the absolute $M^{\star}_g(z=0)$ magnitudes for
the LRGs as in \citet{eisenstein2001}. We examined the photometric errors of the LRGs and deleted the galaxies
brighter than $M^{\star}_g= -23.4$ from the sample to keep the magnitude errors small. In total, our sample includes
170423 LRGs up to the redshift $z=0.6$ (the $k+e$-correction table in \citet{eisenstein2001} stops at this redshift).
It is worth mentioning that the LRG sample is approximately volume-limited (its number density is almost constant)
between the distances from 400 \mh to 1000 \mhp.

We fix the fiducial comoving distance $d_0$ at 435.6 \mh ($z_0=0.15$). The galaxies closer than that are fainter
and are ``not officially'' LRGs \citep{eisenstein2001}. By many properties they are yet similar to LRGs and we need
these to be able to compare the main and LRG superclusters in the volume where the two galaxy samples overlap.

\begin{table*}
\begin{tiny}
\begin{center}
    \caption{\footnotesize Properties of the galaxy samples and of the density fields.}
    \begin{tabular}{rrrrrrrrrrr}
        \hline
        \hline
        Sample & $N_{\mathrm{gal}}$ & $N_\mathrm{groups}$ & $\ell_{\mathrm{mean}}$ & $V_{\mathrm{mask}}$ & $\Delta$ &
        $a$ & $d_{\mathrm{min}}$ \ldots $d_{\mathrm{max}}$ & $z_\mathrm{min}$ \ldots $z_\mathrm{max}$ \\
         & & & $\frac{10^{10} h^{-2} L_\odot}{(\vmh)^3}$ & $(\vgh)^{3}$ & \mh & \mh & \mh & \mh \\
        \hline
        Main        & 583362    & 78800 &   1.526$\cdot10^{-2}$ &   0.132 &     1 &     8     & 55 \ldots 565    & 0.02
        \ldots 0.2 \\
        LRG         & 170423    & -     &   8.148$\cdot10^{-4}$ &   1.789 &     2 &     16    & 60.5 \ldots 1346.4 &
        0.02 \ldots 0.5 \\
        Millennium  & 1039919   & -     &   1.304$\cdot10^{-2}$ &   0.125 &     1 &     8     & -     & - \\
        \hline
    \label{tab:dens_table}
    \end{tabular}\\
\end{center}
\tablefoot{
$\ell_{\mathrm{mean}}$ -- mean density; $V_{\mathrm{mask}}$ -- sample mask volume; $\Delta$ -- grid cell
length; $a$ -- smoothing length; $d$, $z$ -- distance and redshift limits.
}
\end{tiny}
\end{table*}

\begin{figure}
    \input{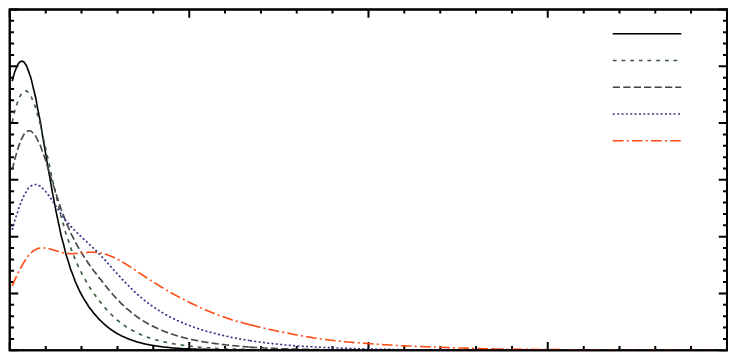}\\
    \input{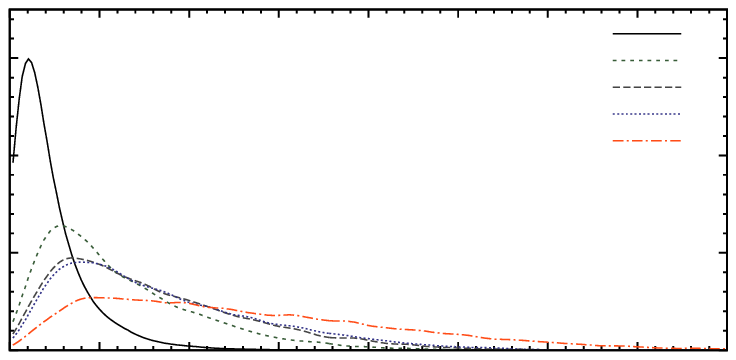}
    \caption{\footnotesize The nearest neighbour distance distribution of the SDSS main ({\it upper panel}) and LRG
    sample ({\it lower panel}) galaxies. The distribution is shown for various distance intervals.}
    \label{fig:nnb}
\end{figure}

\begin{figure}
    \input{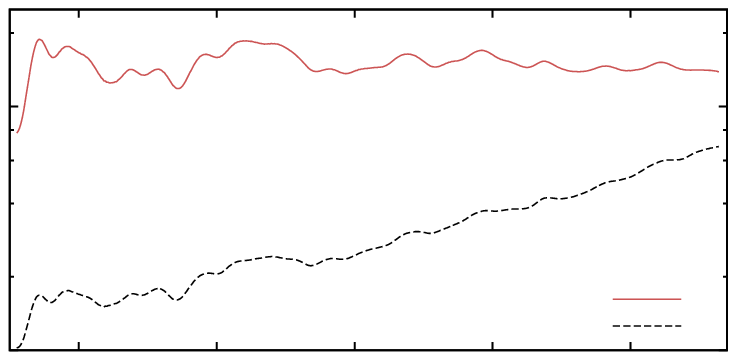}\\
    \input{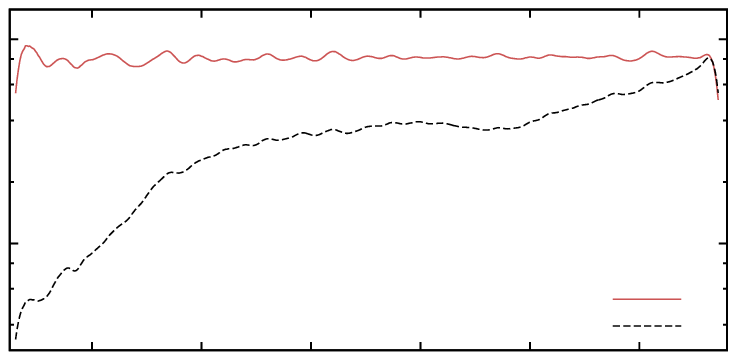}
    \caption{\footnotesize The distance dependence of the average density and of the standard deviations for the main
    ({\it upper panel}) and LRG samples ({\it lower panel}).}
    \label{fig:sigmadist}
\end{figure}

\subsection{The Millennium galaxy catalogue}
We chose a catalogue by \citet{bower2006} that is an implementation of the Durham semi-analytic galaxy
formation model on the Millennium Simulation by the Virgo Consortium \citep{springel2005}. The catalogue is
available from the Millennium database at the German Astrophysical Virtual
Observatory\footnote{\url{http://www.g-vo.org/www/Products/MillenniumDatabases}}. A subsample of about one million
galaxies was selected by the condition $M_r > -20.25$. This yielded a sample with almost the same number density of
galaxies as that of the SDSS main sample (from 125 to 400 \mhp). We calculate the absolute luminosities for galaxies by
taking $M_{\odot}= 4.49$ and using the SDSS $r$ magnitudes (Vega) presented in the catalogue. This sample serves as a
volume-limited test catalogue to study the performance of the supercluster finding algorithm.

\section{Density fields and superclusters}
\label{sec:df_and_scl}

\subsection{Density fields}
\label{sec:df_prop}

We chose the smoothing width of $a = 8$ \mh for the SDSS main sample. The choice of the kernel width is somewhat
arbitrary, but an argument can be made that the scale has to correspond to the size of the structures we are searching
for. For example, the kernel should be considerably wider than the diameters of galaxy clusters, which are a few
megaparsecs. Also, we wish to be able to detect structures at large distances, where galaxies are sparser. We assume
that the density field ties the galaxies together if these are separated by $2a$. Figure \ref{fig:nnb} shows the nearest
neighbour distributions for different distance intervals. We see that for the SDSS main sample the scale $a = 8$ \mh is
comfortably large enough to group galaxies together even at far distances, and a slightly narrower kernel would also
be sufficient. Historically, $a = 8$ \mh has been used in previous supercluster catalogues \citep{einasto2007} and in
other supercluster studies \citep[as in a more recent paper][]{costa2011}. As shown by \citet{costa2011}, the
density field method is actually not very sensitive to the choice of kernel width.

\begin{figure*}
    \input{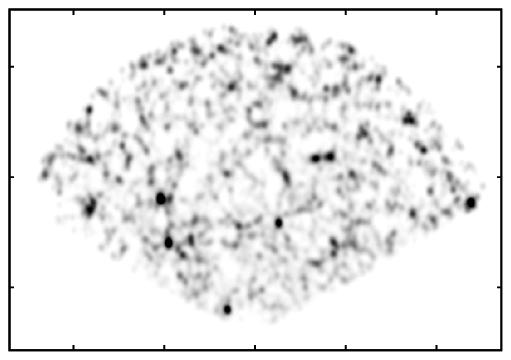}
    \input{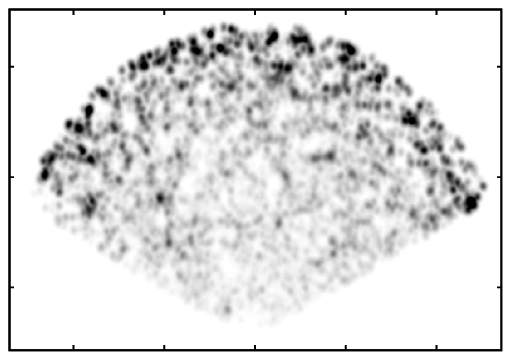}
    \input{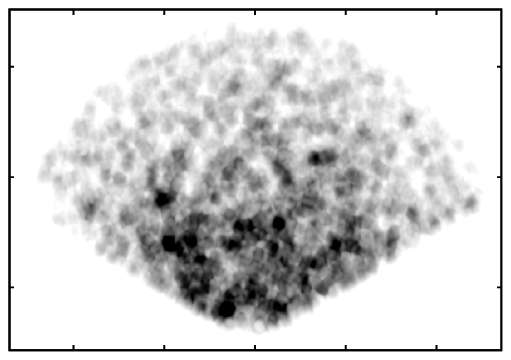}
    \caption{\footnotesize A spatial slice of the luminosity density field for the main sample ({\it left panel}), the
    standard deviation field of the luminosity density ({\it middle panel}), and the signal-to-noise ratio field ({\it
    right panel}). The slice has a thickness of 1 \mh and is located at $z = 33$ \mh (Eq.~\ref{eq:xyz}).}
    \label{fig:snsigmaim}
\end{figure*}

\begin{figure*}
    \input{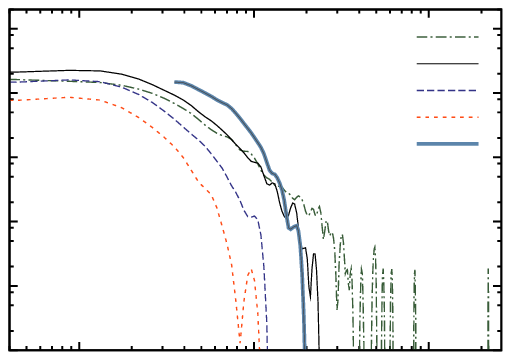}
    \input{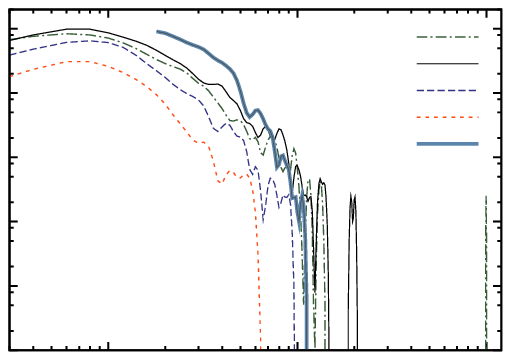}
    \input{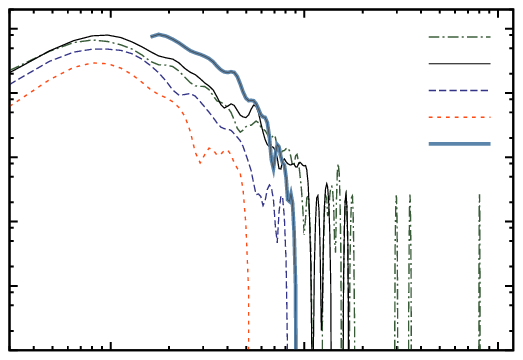}\\
    \input{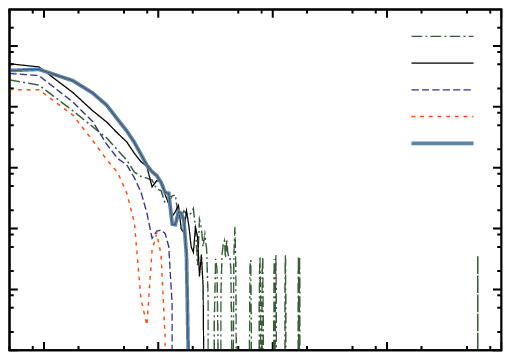}
    \input{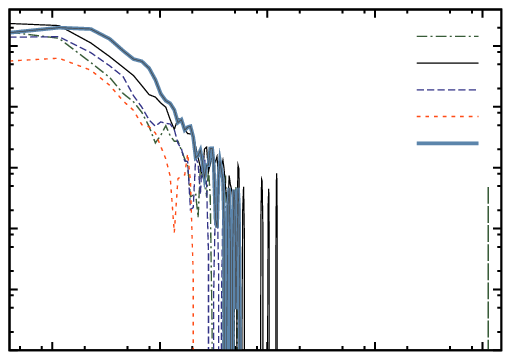}
    \input{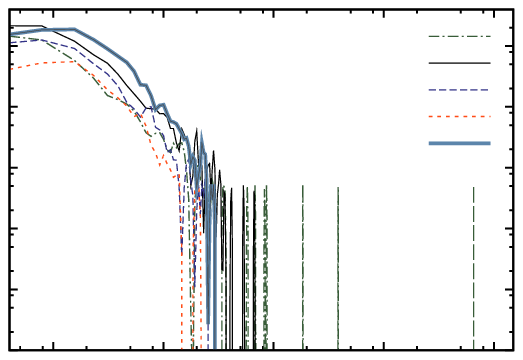}
    \caption{\footnotesize The supercluster diameter ({\it top row}) and its weighted luminosity ({\it bottom row}) in
    the LRG, main, and Millennium samples ({\it from left to right}). Different lines correspond to different density
    thresholds.}
    \label{fig:dff}
    \label{fig:lff}
\end{figure*}

We first employ the sky projection mask (Fig.~\ref{fig:mask}) used in \citet{martinez2009} and then set the lower and
higher limits for the distance. We do not need to use the more precise mask (e.g., the ``mangle'' mask provided by the
NYU VAGC) because we are searching for structures of much larger dimensions. The angular diameter of the kernel at the
far end of the sample is much larger (1.6 degrees for the main sample and 1.3 degrees for the LRGs) than these of the
multitude of small holes inside the SDSS survey mask (with diameters less than an arcminute). The main sample density
field mask is limited within the distances 55 to 565 \mhp. The distance limits here and also in case of the LRGs are
chosen to avoid the distant incomplete regions.

We chose the kernel width for the SDSS LRG sample as $a = 16$ \mh, twice the scale of kernel used for the main
sample, since the LRG sample is sparser. Figure~\ref{fig:nnb} demonstrates that most LRGs have at least one
neighbour at distances up to $2a = 32$ \mhp. The density field of the Millennium sample is calculated with $a = 8$ \mh
kernel width. The mask is a cube with side length of 500 \mhp. Properties of the luminosity density fields for all three
samples are given in Table \ref{tab:dens_table}.

\subsection{Uncertainty analysis}

Following the procedure described in Appendix~\ref{app:uncert} we created 100 realisations of both the main and LRG
samples by randomly shifting galaxies. The shift scale was 8 \mh for the main and 16 \mh for the LRG sample.
Figure~\ref{fig:sigmadist} shows the dependence of $\sigma_\ell$ on distance. We can see the expected rise of
$\sigma_\ell$ with a distance that is mainly caused by the decrease in the galaxy number density. We also see that the
absolute values of the standard deviation are very low when compared to the density. This can be attributed to both
the stability of the large-scale structures and the large smoothing scale for the density fields -- several tens of
galaxies contribute to the density at any point. Example maps of the density, standard deviation, and signal-to-noise
fields spatial slices of the main sample are shown in Fig.~\ref{fig:snsigmaim}. Looking at the images of the standard
deviation field and the signal-to-noise field we can relate them to the observed large-scale structure. Nearby peaks in
the density field stand out also in the signal-to-noise map, but the distant peaks already drown in the noise.

\subsection{Properties of superclusters}
\label{sec:scl_prop}
\subsubsection{Superclusters of the main and LRG samples.}

In this section we describe the general properties of the superclusters and also compare the fixed and
adaptive threshold catalogues. We chose the density difference between the thresholds as $\delta D = 0.1$ (in the
units of the mean density). We compare the diameters and the total weighted luminosities. At this stage we do not
limit our object sample in any way -- it also contains small objects that may consist of only one galaxy and
superclusters located close to the boundary.

\begin{figure}
    \input{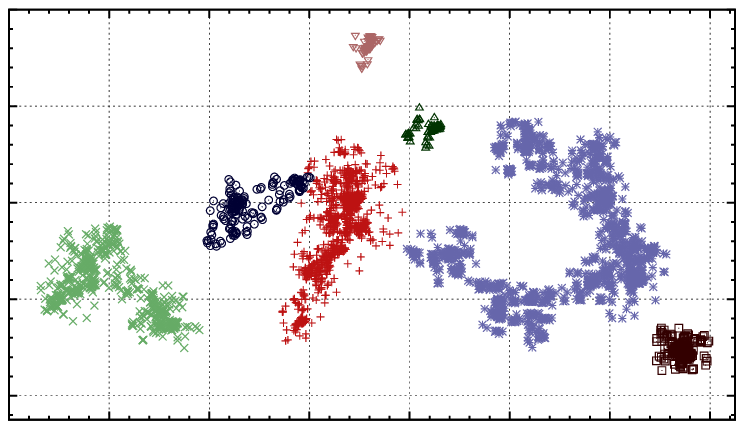}\\
    \input{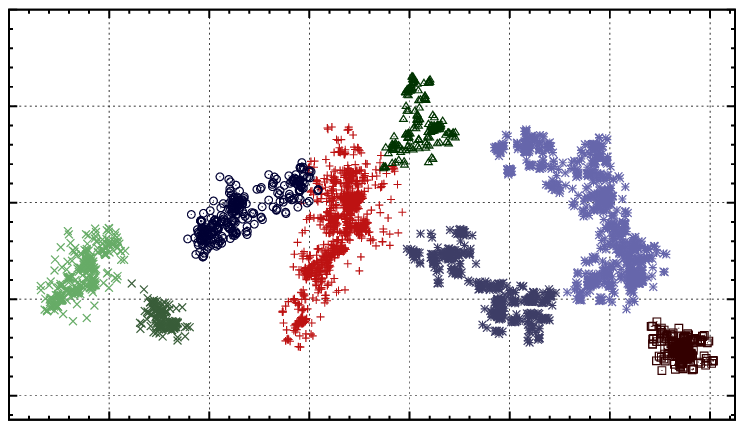}
    \caption{\footnotesize The supercluster SCl~1 and the surrounding superclusters found with different methods for
    assigning the limiting threshold $D$. In the upper panel all superclusters have the same threshold $D_{\mathrm{fix}}
    = 5.0$. The adaptive density levels $D_a$ in lower panel are: SCl~1 -- $D_a = 4.7$; SCl~64 and SCl~378 -- $D_a =
    5.5$; SCl~94 and SCl~368 -- $D_a = 5.3$; SCl~362 -- $D_a = 4.4$; SCl~578 -- $D_a = 4.3$; SCl~1310 -- $D_a = 3.7$.}
    \label{fig:adapfixd}
\end{figure}

Figure~\ref{fig:dff} shows the diameter and weighted luminosity distributions for the main and LRG samples for different
density levels. The shapes of the curves are similar in both samples, with the diameter distribution offering a slightly
clearer picture. For all density thresholds, the maximum of the curves is located at about the same diameter/luminosity
value. The slight dip in the distributions at small and dim objects is caused by our not including any
density field objects that have no galaxies inside. At the high diameter/luminosity wings, the distributions have a
series of maxima, which are characteristic of all density thresholds. This is caused by structures that are
distinctively larger than most of the objects, and they are present even at high density levels ($D = 8.0$). As we move
towards lower density levels ($D = 6.0$), the number of objects increases, while the objects themselves also get larger.
The maxima caused by very large superclusters also become more prominent and at some point separate from the main body
of the distribution, as they begin to include increasing numbers of smaller objects ($D = 4.0$). At the lowest density
threshold, below percolation, there is one enormous structure that extends throughout the sample volume (at $D = 2.0$).

In Fig.~\ref{fig:dff} the distributions for the adaptive catalogues start at the minimum distance limit. They have
higher values than the distributions for the fixed level superclusters because they include contributions from
superclusters at several density
thresholds.

Figure~\ref{fig:adapfixd} presents an example of how the superclusters are affected by the two selection methods, fixed
or adaptive density thresholds. The most noticeable consequence is that the superclusters SCl~64 and SCl~94 in the upper
panel (a fixed density level) have both been broken in two and all their components have thresholds higher than before,
while the superclusters SCl~1, 362, 578 have been defined at lower density levels and SCl~1310 has been assigned a much
lower density level and is considerably larger because of that. The supercluster SCl~1320 does not meet the minimum
diameter criterion ($\diameter \geq 16 \vmh$) and is not included in the catalogue with adaptive thresholds. If an
object fails to qualify as a supercluster, it does not necessarily mean that galaxies belonging to this object are
absent from the catalogue; instead of that, they can belong to some other supercluster at a lower threshold. This
depends on the specific geometry of the supercluster environment.

\begin{figure*}
    \input{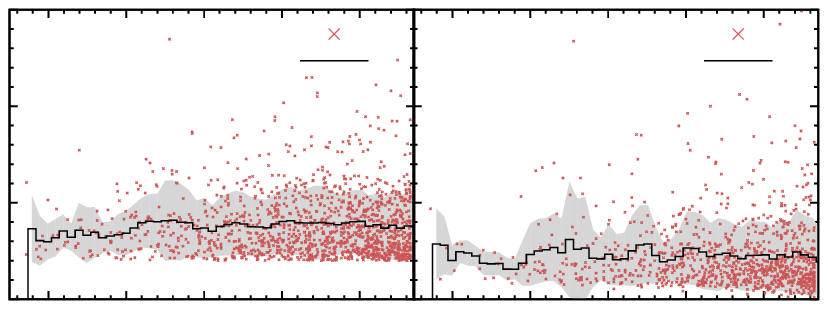}
    \input{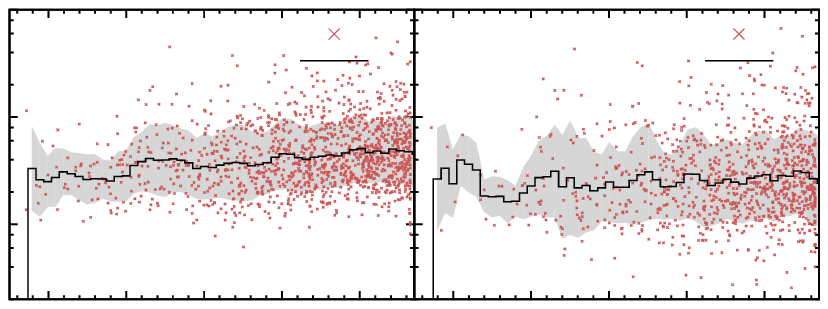}\\
    \input{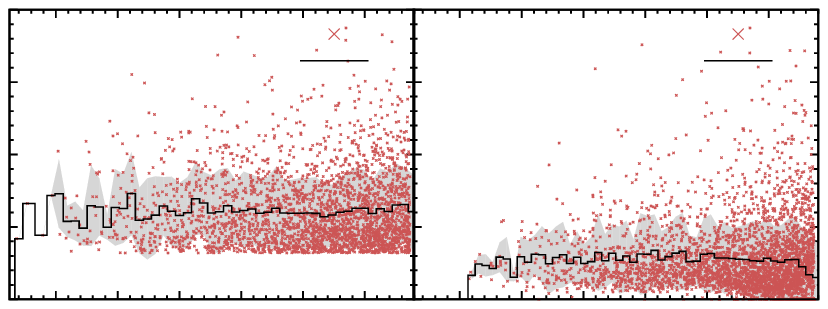}
    \input{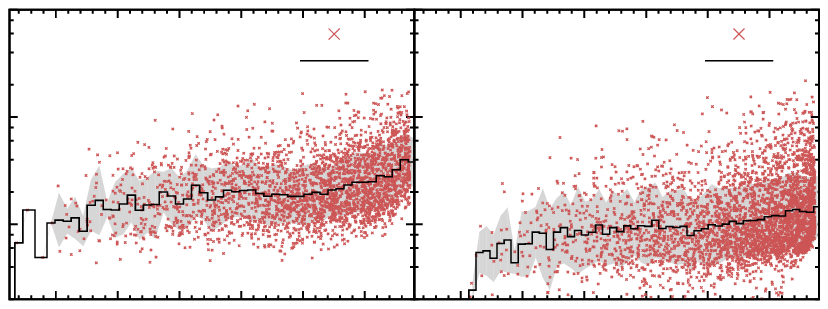}
    \caption{\footnotesize Supercluster diameters ({\it left panels}) and total weighted luminosities ({\it right
    panels}) vs distance for fixed and adaptive thresholds. The main sample is shown on the upper row and LRG on the
    lower. Points mark the diameters and luminosities of individual superclusters; the line is the average with the bin
    widths of 10 \mh for the main and 25 \mh for the LRG sample. The standard deviations in bins are shown with
    grey contours.}
\label{fig:distdep}
\end{figure*}

We have to check whether the supercluster properties depend on distance and how the different density level assignments
work. Figure~\ref{fig:distdep} shows the dependence of the diameter and the weighted luminosity on distance. The extent
of the scatter of diameters and luminosities increases with distance, while the averages remain more or less stable, and
the standard deviations also do not exhibit systematic increase or decrease with the distance. The average diameter is
almost constant for both the main and LRG samples. The barely noticeable downward trend in the fixed threshold
supercluster catalogue is caused by small galaxy groups or even single galaxies, which are bright but do not form
larger structures because of the sparseness of the galaxy sample. The weighted luminosity, however, tends to rise
slightly for the main sample, and in a quite obvious manner for the LRG sample. Together, these graphs suggest that
superclusters with similar dimensions are brighter at large distances, which implies some overweighting.

\begin{figure}
    \input{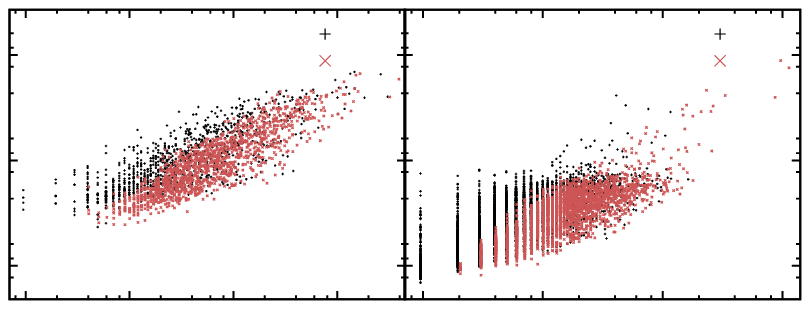}
    \input{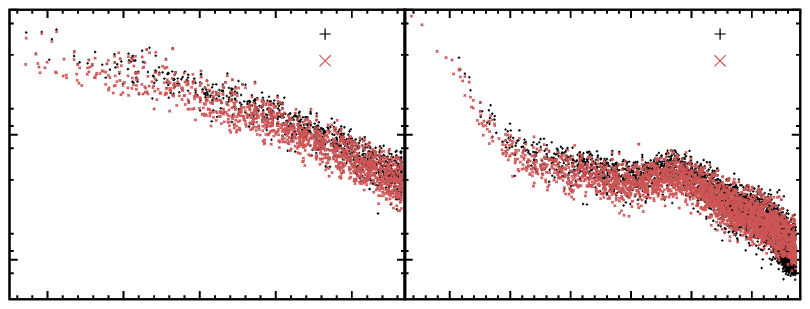}
    \caption{\footnotesize Supercluster confidence estimates vs their richness (the number of member galaxies) ({\it top
    panels}) and vs their distance ({\it bottom panels}). The superclusters defined by a fixed threshold are marked with
    black, and those found using adaptive thresholds with red points.}
    \label{fig:sclsn}
\end{figure}

Figure~\ref{fig:sclsn} shows the dependence of the supercluster confidence estimates on supercluster richness (its
number of galaxies) and distance. The confidence estimates are calculated as in Eq.~(\ref{eq:sn}). Both graphs display
the expected behaviour. The confidence estimates diminish with distance, and richer superclusters also have higher
signal-to-noise ratios. This property can be used to select objects for further studies. Predictably, the confidence
estimates for superclusters in the LRG sample are significantly lower. The confidence estimates depend on the density
threshold, but at lower density levels, more galaxies from the density field regions with higher variance are included.
Because of that, fixed threshold superclusters have higher confidence estimates in Fig.~\ref{fig:sclsn}.

\begin{figure}
    \input{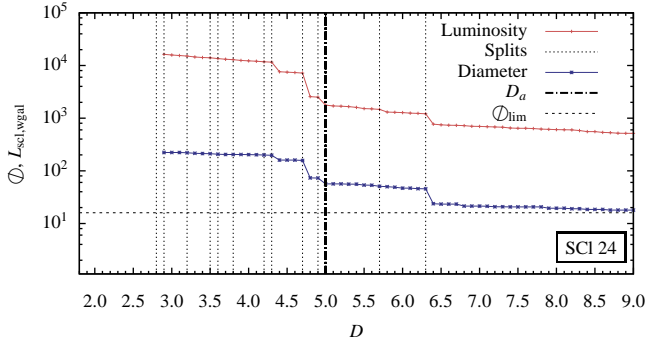}
    \caption{\footnotesize An example of the dependence of the supercluster diameter and the total luminosity on the
    density level $D$. Vertical thin dashed lines show splitting/merger events, and the thick dashed line shows
    the adaptive density threshold. The minimum diameter limit 16 \mh is also shown. The lines begin at the level where
    the object separates from the larger structure.}
\label{fig:sclpropD}
\end{figure}

\begin{figure}
    \input{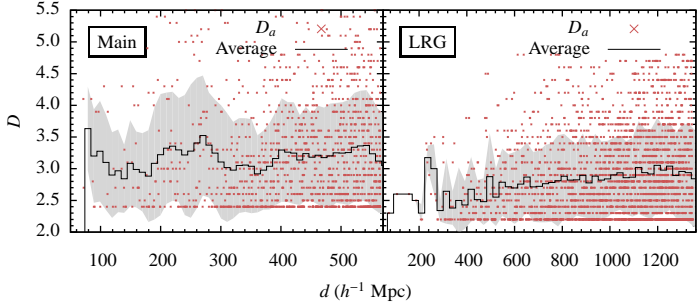}
    \caption{\footnotesize Adaptively assigned supercluster thresholds, the average and the standard deviation vs
    distance.}
    \label{fig:ddist}
\end{figure}

\begin{figure}
    \input{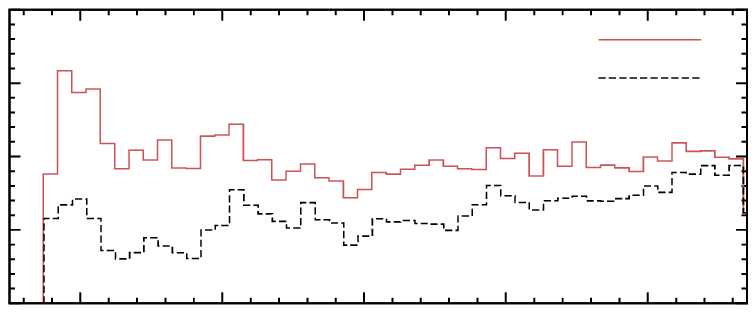}\\
    \input{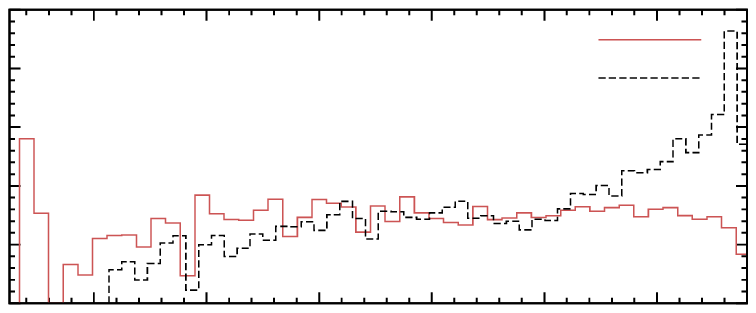}
    \caption{\footnotesize The dependence of number densities on distance for adaptive and fixed threshold
    superclusters. The main sample is shown in the upper panel and the LRG sample in the lower panel.}
    \label{fig:sclndist}
\end{figure}

Next we take a look at structure breakups and adaptively assigned supercluster thresholds. Figure~\ref{fig:mergs} shows
the number of splitting events, the percolation level, and the 95\% limiting density threshold for $D_{\mathrm{scl}}$.
Figure~\ref{fig:sclpropD} gives an example of how supercluster diameters and luminosities change drastically during
mergers when lowering the density level and, while still growing, remain relatively stable in between. Supercluster
SCl~24 in Fig.~ \ref{fig:sclpropD} is a part of the Sloan Great Wall and at densities $D<4.7$ it actually includes all
of the SGW superclusters \citep{maret2010}. For the main sample, density threshold assignments do not show any clear
dependence on distance, while for the LRG sample the adaptively found levels are increasing with distance
(Fig.~\ref{fig:ddist}). The broad peaks, which are visible in Figs.~\ref{fig:distdep}, \ref{fig:ddist}, and
\ref{fig:sclndist} at approximately 250 \mh are caused by the Sloan Great Wall region superclusters.

Selection effects can cause the number density of superclusters to depend on distance. Figure~\ref{fig:sclndist}
demonstrates that using a single density level to define superclusters causes a significant rise in the number of
objects with distance. The reason for this is the Poisson noise that is caused by the increased density contrast because
of the weighting. As mentioned before, we can add the missing luminosity only where we see the galaxies, but not there
where it is actually missing. In contrast, the number density of adaptive threshold catalogue superclusters is
independent of distance.

\subsubsection{Superclusters of the Millennium sample}

We used the Millennium galaxy sample to evaluate the supercluster-finding procedure as applied to an ideal
volume-limited sample. As a consequence, there are no distance-related effects. A separate further study will look more
closely at the differences in object selection and their properties using several simulated flux and volume-limited
samples.

\begin{figure*}
    \input{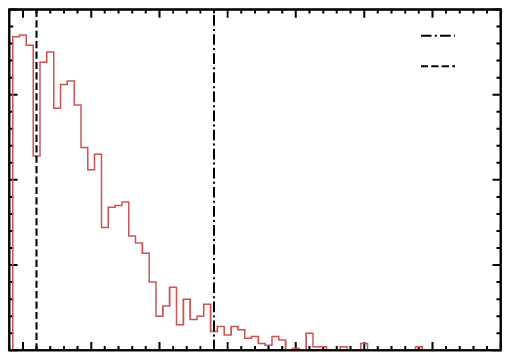}
    \input{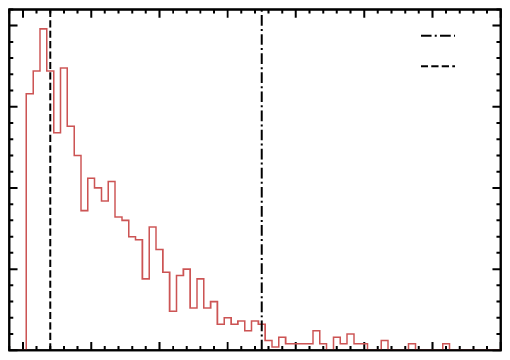}
    \input{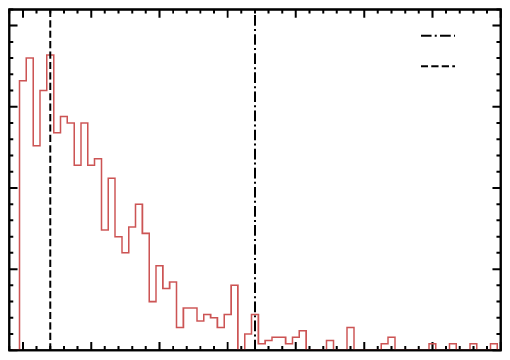}
    \caption{\footnotesize The number of structure breaks (anti-mergers) per density threshold. Vertical lines denote
the percolation and the upper limit densities. From the left: LRG, main, Millennium.}
\label{fig:mergs}
\end{figure*}

Figure~\ref{fig:dff} allows us to compare the supercluster diameter and weighted luminosity distributions for the main
sample to those for the Millennium sample. The distributions are strikingly similar, with the main sample containing
almost the same amount of superclusters. The shape of the Millennium distributions at the density level $D = 8.0$ also
indicates the presence of large Sloan Great Wall-like structures. Still, in contrast to the main sample, at the lowest
density threshold there are also other large structures besides the one that has percolated. Also, the adaptive
level superclusters appear to be slightly smaller than those in the main sample, which does not contradict earlier
similar findings \citep{einasto2006}. The number of structure splits versus the density threshold graphs for the
Millennium and the main sample are virtually indistinguishable. They also share the percolation threshold and the 95\%
maximum density level differs by only one $\delta D$ (Fig.~\ref{fig:mergs}).

The summary of the properties of the supercluster catalogues for all three samples is given in
Table~\ref{tab:scl_table}. Both the main and LRG samples have most superclusters at the same threshold $D = 3.0$, with
1566 and 4780 objects, accordingly (the volume of the LRG sample is about 14 times larger than that of the main sample).
There is only one major difference with the Millennium sample: it has most superclusters, 1316 at a slightly higher
threshold $D = 3.3$, than the observational samples ($D = 3.0$). We find that significantly more galaxies belong to
superclusters in the adaptive catalogue. For main and Millennium samples, the percentage rises from about 15\% to
more than a quarter of the galaxies, and in the LRG sample about 80\% of the galaxies belong to superclusters in the
adaptive threshold catalogue.

Comparison with the volume-limited Millennium sample shows that our supercluster algorithms generally work well and
that we have avoided the selection problems inherent to magnitude-limited samples.

\subsubsection{Large-scale variations in the SDSS main sample}
\begin{figure}
    \input{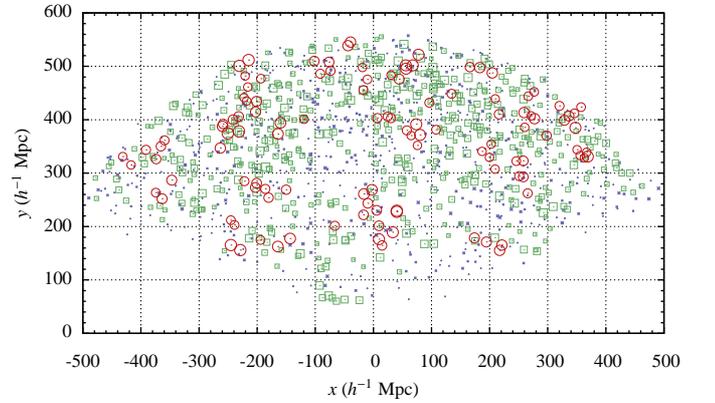}
    \caption{\footnotesize The SDSS main sample supercluster map. Different symbol types and sizes show the density
    threshold levels used to delineate the superclusters; blue points: $D \leq 3.0$, green squares: $4.0 < D \leq
    4.5$, and red rings: $D > 4.5$. The map is the 2-D projection of the whole supercluster sample. Substantial
    differences in the levels can be seen e.g., in the regions around (-60,300) and at the Sloan Great Wall region at
    (0,220).}
    \label{fig:SDSSmap}
\end{figure}

If we look at the positions and density levels of the adaptive-threshold supercluster map of the SDSS main sample
(Fig.~\ref{fig:SDSSmap}), we see that there are strong variations in the supercluster thresholds depending on the
region where they are located. The threshold level needed to define a supercluster is tightly correlated with the
overall mean density. The spatial scale of these variations is about 200--300 \mhp. One can discern the dominant
supercluster plane \citep{maret1997} and system of large voids behind it \citep[described in detail
by][]{platen2009phd}. The fact that these variations are not lost in the projection (we show the 2-D projection of the
full Legacy volume) shows that they are really huge. The reason for these variations is presently unclear, so we leave
their quantification and study for the future.

\begin{table*}
\begin{tiny}
\begin{center}
    \caption{\footnotesize Supercluster catalogue properties.}
    \begin{tabular}{rrrrrrrrr} 
        \hline
        \hline
        (1)&(2)&(3)&(4)&(5)&(6)&(7)&(8)\\
        \hline 
        Sample & $N_{\mathrm{scl}}$ & $n_{\mathrm{scl}}$& $f_\mathrm{edge}$ & $f_{\mathrm{gal}\in\mathrm{scl}}$ &
        $D_{\mathrm{fix}}$ &
        $D(N_{\mathrm{max}})$ & $n_{\mathrm{scl}}(D(N_{\mathrm{max}}))$\\
         & & ($h^{-1}$ Gpc)$^{-3}$ & & & $\ell_{\mathrm{mean}}$ & $\ell_{\mathrm{mean}}$ & ($h^{-1}$ Gpc)$^{-3}$\\
        \hline
        Fixed threshold\\
        Main        & 982   & 7432 & 0.184 & 0.138 & 5.0 & 3.0 & 11852  \\ 
        LRG         & 3761  & 2101 & 0.151 & 0.184 & 4.4 & 3.0 & 2671   \\ 
        Millennium  & 844   & 6752 & 0.147 & 0.153 & 5.0 & 3.3 & 10528  \\ 
        \hline
        (1)&(2)&(3)&(4)&(8)&(9)&(10)&(11)\\
        \hline 
        Sample & $N_{\mathrm{scl}}$ & $n_{\mathrm{scl}}$& $f_\mathrm{edge}$ & $f_{\mathrm{gal}\in\mathrm{scl}}$ &
        $D_{\mathrm{lim}}$ &
        $D_{\mathrm{perc}}$ & $\diameter_{\mathrm{lim}}$ \\
         & & ($h^{-1}$ Gpc)$^{-3}$ & & & $\ell_{\mathrm{mean}}$ & $\ell_{\mathrm{mean}}$ & \mh \\
        \hline
        Adaptive threshold\\
        Main        & 1313 & 9937 & 0.225 & 0.267 & 5.5 & 2.4 & 16 \\ 
        LRG         & 2701 & 1509 & 0.153 & 0.822 & 4.8 & 2.2 & 32 \\ 
        Millennium  & 1214 & 9712 & 0.194 & 0.282 & 5.4 & 2.4 & 16 \\ 
        \hline
    \label{tab:scl_table}
    \end{tabular}\\
\end{center}
\tablefoot{
Columns in the Table:
1: sample name and threshold assigning method;
2: the number of superclusters;
3: the number density of superclusters;
4: the fraction of superclusters close to the sample edge;
5: the fraction of galaxies in superclusters;
6: the fixed threshold value;
7: the density threshold with most objects;
8: the number density of objects for the threshold $D(N_{\mathrm{max}})$;
9: the maximum allowed value for $D_a$;
10: the percolation threshold;
11: the minimum allowed supercluster diameter.
}
\end{tiny}
\end{table*}

\section{Conclusions and discussion}

Superclusters are the elements of the overall large-scale structure, the ``LEGO pieces'' of the Universe. As such, they
describe the whole cosmic web of galaxies. They are also the largest objects of that web, and, although they are not
gravitationally bound, in the future they may become bound isolated structures, the real ``island Universes''
\citep{araya2009}.

Developing supercluster catalogues is useful for future, and sometimes unexpected, applications. The study of quasar
environments \citep{lietzen2009} is a natural application of the multilevel supercluster catalogue, by aiding the
uniform description of the overall matter density field. Searches for specific directions that are promising for
observations is another example of where supercluster catalogues are indispensable; for example, a search for the
elusive warm-hot intergalactic medium (WHIM) can be more effective with prior knowledge of the structures that are
theoretically associated with the WHIM \citep{fang2010}. And the identification of the Planck SZ source, mentioned in
the introduction, is a perfect example of an unexpected development.

The main result of this work is a set of supercluster catalogues, based on the SDSS DR7 galaxy data, for the main and
the LRG samples. The catalogues are public. We define superclusters, first for different mean density thresholds, and
then for adaptive density thresholds that are different for each supercluster.

It is possible to create almost selection-free samples of superclusters from flux-limited catalogues. We studied the
supercluster properties and found little dependence on the distance. We also compared the SDSS superclusters with the
superclusters based on the Millennium galaxies, which were built using the same algorithms, and the supercluster samples
have very similar properties.

While the LRG sample is very sparse and the number density of superclusters in its volume is much lower than for the
main sample, one can still construct a supercluster sample with comparable properties.

When previous supercluster catalogues were based on fixed density levels (or nearest neighbour distances), we feel that
the multiscaling (multi-threshold) approach is essential for defining the supercluster environment. The multi-level
catalogues are useful for studying the overall density field, but for following individual superclusters, their
structure, and their evolution, the adaptive threshold algorithm produces the best superclusters. With the full fixed
threshold supercluster data set it is possible to create new adaptive threshold catalogues using alternative sets of
limiting parameters. The adaptive threshold supercluster definition procedure permits more galaxies to be included in
more superclusters, while also suppressing the selection effects. It allows us to generate practically volume-limited
supercluster samples. In the LRG sample, the vast majority of galaxies are enclosed in superclusters. This is natural
since LRGs are bright galaxies presumably residing in the cores of large galaxy groups, which in turn are very likely to
be situated in superclusters \citep{einasto2003}.

Galaxy superclusters are fairly well-defined systems. With the increasing density level, the supercluster sizes change
radically with structure breaks, but are relatively stable in between, because they do not acquire or lose many galaxies
while changing the density level. An important point is that at present, the number of known superclusters is
small (especially the number of very large superclusters), which makes it possible to study them individually by looking
at every one of them and correcting the possible glitches in their delineation.

There are certainly problems that remain unresolved at the moment. There is the question of boundary effects, for one
using a fixed distance from the sample edge to limit the supercluster sample, as is sometimes done, is not entirely
justified. First, it removes a large fraction of galaxies from the present samples. Second, many of the large
superclusters (e.g., SCl~126 of the Sloan Great Wall \citep{maret2001}) are touching the SDSS mask edges, but are even
so the largest between the superclusters. In fact, most of the nearby superclusters are incomplete because of the
cone-like shape of the survey. Thus we also included such superclusters, and marked if those that were affected by the
sample borders. It is already the decision of the catalogue users how they take that mark into account. Superclusters
from the LRG sample show clear selection effects at the outer border of the sample volume. This is caused by the low
number density and strong luminosity weighting. An unexpected result is that there is an overall density variation, and
the variation of supercluster properties, on very large scales (about 200 \mhp), in the SDSS Legacy sample volume. We
will discuss that in detail in the next paper.

\begin{acknowledgements}
The catalogue owes much to fruitful discussions with Vicent Mart{\'\i}nez (Valencia) and with the members of the
cosmology group at Tartu Observatory.

Funding for the Sloan Digital Sky Survey (SDSS) and SDSS-II has come from the National Science Foundation, the U.S.
Department of Energy, the National Aeronautics and Space Administration, the Japanese Monbukagakusho, the Max Planck
Society, and the Higher Education Funding Council for England.  The SDSS Web site is http://www.sdss.org/. The SDSS is
managed by the Astrophysical Research Consortium (ARC) for the Participating Institutions.  The Participating
Institutions include the American Museum of Natural History, Astrophysical Institute Potsdam, University of Basel,
University of Cambridge, Case Western Reserve University, The University of Chicago, Drexel University, Fermilab, the
Institute for Advanced Study, the Japan Participation Group, The Johns Hopkins University, the Joint Institute for
Nuclear Astrophysics, the Kavli Institute for Particle Astrophysics and Cosmology, the Korean Scientist Group, the
Chinese Academy of Sciences (LAMOST), Los Alamos National Laboratory, the Max-Planck-Institute for Astronomy (MPIA), the
Max-Planck-Institute for Astrophysics (MPA), New Mexico State University, Ohio State University, University of
Pittsburgh, University of Portsmouth, Princeton University, the United States Naval Observatory, and the University of
Washington. The Millennium Simulation databases used in this paper and the web application providing online access to
them were constructed as part of the activities of the German Astrophysical Virtual Observatory.

We acknowledge the Estonian Science Foundation for support under grants 8005, 7765, and MJD272, the Estonian Ministry
for Education and Science support by grants SF0060067s08, and also the support by the Center of Excellence of
Dark Matter in (Astro)particle Physics and Cosmology (TK120). Computations for the catalogues were carried out at the
High Performance Computing Centre, University of Tartu.
\end{acknowledgements}

\bibliography{dr7cat}{}
\bibliographystyle{aa}

\begin{appendix}
\section{Kernel density estimates}
\label{app:kern}

As superclusters are searched for as regions with the luminosity density over a certain threshold in a compact region of
space, we have to convert the spatial positions of galaxies into a luminosity density field. The standard approach is to
assume a Cox model for the galaxy distribution, where the galaxies are distributed in space according to a inhomogeneous
point process with the intensity $\rho(\mathbf{r})$ determined by an underlying random field \citep[see,
e.g.][]{martinez2003}. The best way to estimate this intensity is by a kernel sum \citep[][sect. 8.3.2]{davison1997}:
\begin{equation}
    \rho(\mathbf{r}) = \frac{1}{a^3}\sum_{i=1}^N K\left( \frac{\mathbf{r} - \mathbf{r}_i}{a}\right),
    \label{eq:dens}
\end{equation}
where the sum is over all $N$ data points, $\mathbf{r}_i$ are the coordinates, $K(\cdot)$ is the kernel,
and $a$ the smoothing scale. As we estimate luminosities, we multiply kernel amplitudes by weighted galaxy
luminosities $L_{\mathrm{gal,w}}$ and calculate the luminosity density field as
\begin{equation}
    \ell(\mathbf{r}) = \frac{1}{a^3}\sum_\mathrm{gal} K\left( \frac{\mathbf{r} - \mathbf{r}_{\mathrm{gal}}}{a}\right)
    L_{\mathrm{gal,w}},
    \label{eq:lum_dens}
\end{equation}

\begin{figure}
    \input{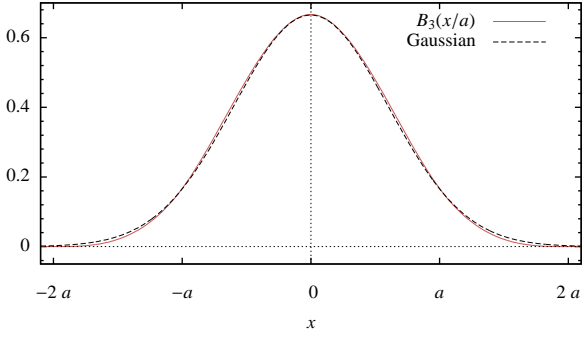}
    \caption{\footnotesize The shape of the kernel $B_3(x)$. Solid line -- the $B_3(x)$ kernel; dashed line -- a
    Gaussian with $\sigma=0.6$.}
    \label{fig:tuum}
\end{figure}

The kernels $K(\cdot)$ are required to be distributions, positive everywhere and integrating to unity; in our case,
\begin{equation}
    \int K(\mathbf{y})d^3y=1.
    \label{eq:kern}
\end{equation}
Good kernels for calculating densities on a spatial grid are the box splines $B_J$. They are local and they are
interpolating on a grid:
\begin{equation}
    \sum_i B_J \left(x-i \right) = 1,
    \label{eq:sum}
\end{equation}
for any $x$ and a small number of indices that give non-zero values for $B_J(x)$. To create our density fields we use
the popular $B_3$ spline function:
\begin{equation}
    B_3(x) = \frac{|x-2|^3 - 4|x-1|^3 + 6|x|^3 - 4|x+1|^3 + |x+2|^3}{12}.
\end{equation}
This function differs from zero only in the interval $x\in(-2,2)$, meaning that the sum in (\ref{eq:sum}) only includes
values of $B_3(x)$ at four consecutive arguments $x\in(-2,2)$ that differ by 1. In practice, we calculate the kernel sum
(\ref{eq:kern}) on a grid. Let the grid step be $\Delta<a$, and $a=k\Delta$, where $k>1$ is an integer. Then the sum
over the grid
\begin{equation}
    \sum_i B_3 \left(\frac{x-i\Delta}{a}\right) = k,
    \label{eq:scaledsum}
\end{equation}
because it consists of $k$ groups of four values of $B_3(\cdot)$ at consecutive arguments, differing by 1. Thus, the
kernel
\begin{equation}
    K_B^{(1)}(x/a;\Delta) = \frac{\Delta}{a}B_3(x/a)
\end{equation}
differs from zero only in the interval $x\in[-2a,2a]$ (Fig.~\ref{fig:tuum}) and preserves the interpolation property
exactly for all values of $a$ and $\Delta$, where the ratio $a/\Delta$ is an integer (also, the error is very small even
if this ratio is not an integer, but $a$ is at least several times larger than $\Delta$). The three-dimensional kernel
$K_B^{(3)}$ is given by a direct product of three one-dimensional kernels:
\begin{eqnarray}
    K_B^{(3)}(\mathbf{r}/a;\Delta) &\equiv& K_B^{(1)}(x/a;\Delta) K_B^{(1)}(y/a;\Delta) K_B^{(1)}(z/a;\Delta)\\ &=&
    \left(\frac{\Delta}{a}\right)^3 B_3(x/a)B_3(y/a)B_3(z/a),
    \label{eq:3dkern}
\end{eqnarray}
where $\mathbf{r} \equiv \{x,y,z\}$. Although this is a direct product, it is practically isotropic
\citep{saar2009}. This can be seen already from the fact that it is very close to a Gaussian with a mean zero and
$\sigma=0.6$ (Fig.~\ref{fig:tuum}), and the direct product of one-dimensional Gaussians is exactly isotropic.

\section{Error analysis of the density field}
\label{app:uncert}
To characterise the errors of our density field estimates we have to choose the statistical model for
the galaxy distribution. The most popular model used for the statistics of the spatial distribution of galaxies in the
Universe is the ``Poisson model'' \citep{peebles1980}, an inhomogeneous Poisson point process where the local intensity
of the process is defined by the amplitude of the underlying realisation of a random field. In statistics it is called
the Cox random process, see an introduction and examples in \citet{martinez2003} and \citet{illian2008}. In cosmology,
the random fields used are usually Gaussian or log-Gaussian fields.

As for any statistical model, it has been postulated to describe the galaxy distribution, and its success in
applications describing the statistical properties of that distribution tends to support it. For example, this model was
used to develop methods for estimating the two-point correlation function \citep{hamilton1993} and the power spectrum of
the galaxy distribution \citep{tegmark1998}. These methods have been extensively used to study the galaxy distribution.
The same model serves as the basis for a maximum-likelihood approach to recover the large-scale cosmological density
field by \citet{kitaura2010}.

We use the kernel method to estimate the intensity of our Cox process. A popular procedure for estimating the
uncertainties of kernel-based intensity estimates for inhomogeneous Poisson processes is bootstrap \citep[see,
e.g.,][sect. 8.3.2]{davison1997}. Because the kernel used in estimating the intensity in Eq.~\ref{eq:dens} is compact,
there is only a finite and, in practice, a relatively small number of members in this sum. Bootstrap is used to estimate
the sample errors (discreteness errors) caused by the discrete sampling. As stressed by \citet{silverman1987}, bootstrap
consists of two separate elements. Let our sample be $(X_1,\dots,X_n)$. First, to estimate the discreteness error caused
by the finite sample size $n$ of the sample parameter $\theta(X)$ that we are estimating, we use the sampling method,
drawing a large number of samples of size $n$ from the (integral) population distribution function $F(X)$. Technically,
it is the simplest method for generating random numbers with a given distribution -- select $n$ uniform random numbers
$U_i$ in the interval (0,1) and select the sample value $X$ given by $F(X)=U$ for each $U$. The other element of
bootstrap is to assume that the  population distribution $F$ can be approximated by the empirical distribution function
$F_n$ defined by all the $n$ observed values $(X_1,\dots,X_n$) that form the sample. If all $X_i$-s are i.i.d.
(independent and identically distributed), this function can be defined as a step function with increments $1/n$ at
every $X^\star_j$, where $(X^\star_1,\dots,X^\star_j)$ is an ordered growing sequence of the original sample values
$X_i$.  If we select from $F_n$, any bootstrap sample consists of the values of the original sample, selected from the
original sample randomly with replacement \citep{efron1993}. Using the values of $\theta$ for all bootstrap samples, we
can find the sampling errors (usually the bias and the variance) of the parameter $\theta$. These are the bootstrap
error estimates we search for.

Theorems that prove the effectiveness of the bootstrap error estimates are usually proved for the case where both
the sample size and the number of the bootstrap samples approach infinity \citep{shao1995}, and for a finite sample size
simulations are used. In our case, the sample that defines the intensity estimate for a given point in space consists
of these galaxies, the positions of which are within the kernel volume around that point. The usual rule of thumb says
that bootstrap error estimates may be considered reliable when the sample size is more than 30 (even sizes as small as
14 have been used in simulation examples, as in \citet{efron1993} and \citet{silverman1987}). Our kernel volumes
include, on average, 150 galaxies in the case of the SDSS main sample and 25 galaxies for the LRG sample. At the density
levels where we define our superclusters, $(D\approx5)$, the corresponding numbers are 750 and 125, so our error
estimates should be reliable enough.

For the inhomogeneous Poisson process, where the points $\mathbf{X}_i$ are identically and independently distributed
with the locally defined intensity $\lambda$, bootstrap can be used to estimate the errors of the kernel estimate
(Eq.~\ref{eq:dens}). In practice, for intensity estimation a bootstrap version that is called a smoothed bootstrap is
used. This is a version of parametric bootstrap, using, instead of the empirical distribution function, its smoothed
version. \citet{silverman1987} demonstrated that it is more effective in estimating the variance of intensity as the
standard bootstrap. To use that, in practice, we generate bootstrap samples of the same size as the original sample,
selecting the galaxies from our sample randomly with replacement, as usual in bootstrap, but give the selected galaxies
random displacements. As explained in \citet{davison1997} and \citet{silverman1987}, the random spatial displacements
are required to have the probability density of the same form as the kernel function, but it is useful to undersmooth,
using the kernel for the displacements  that is narrower than the kernel used for calculating the intensity estimates.
We undersmooth by a factor of two. As our grid size is huge (about $10^8$), we use 100 bootstrap samples for each grid
vertex. This number has been found to be large enough to estimate the sample variance, based on simulation studies
\citep{efron1993}.

Another point that has to be taken care of when estimating global (population) statistics as correlation
functions or power spectra (spectral densities) of Cox processes is to account for the difference of the measured
statistic for the specific realisation of the random field and for the random field as a whole (the so-called cosmic
noise problem, see, e.g., \citet{szapudi1996}, \citet[][p. 522]{peacock1999}). The discreteness errors, estimated by
bootstrap, and the realisation variance combine in a subtle way \citep{cohn2006}. In our case, fortunately, the spatial
density -- the intensity of the Cox process that we are estimating (the geography of the large-scale structure) is
exactly the underlying random realisation itself -- we are measuring the cosmic noise, so are not interested in the
mean density of the Universe. The only errors our intensity estimates have are discreteness errors, and these can be
estimated by bootstrap.

We select the galaxies for the bootstrap samples, together with their measured luminosities, and we consider galaxy
distribution as a marked Cox process, with luminosities as marks. If we could statistically model the luminosity
distribution among galaxies as random \citep[a random marks model, see, e.g.,][]{illian2008}, we could build bootstrap
samples by randomly relabelling galaxies, choosing their luminosities as in the usual bootstrap from the luminosities of
the sample galaxies inside the kernel volume. This, however, would not be right, as galaxies are well known to be
segregated by luminosity -- more luminous galaxies populate regions of higher number density of galaxies
\citep{hamilton1988, girardi2003}. We chose another way and tried modelling the luminosity errors. These consist of a
small error of the luminosity weights, generated by the errors of the luminosity function, and an error in modelling the
evolution correction \citep{blanton03b}. We tested the effect of these errors by selecting them randomly from the
observed distributions, compared the intensity estimates with modified luminosities and with fixed luminosities, and
found no significant differences. As the luminosity errors were much smaller than the deviations of the intensity
estimates generated by bootstrap, the discreteness errors, and we did not find a good statistical model to describe
them, we ignored these errors.

After calculating the positions for the galaxies of a bootstrap sample, we find a new  intensity estimate. We repeated
the procedure a number of times (for this paper, we generated 100 bootstrap samples for every grid point where we
estimated the intensity) and found the standard deviation for the intensity $\sigma_\ell$ for each grid vertex as
\begin{equation}
    \sigma_\ell = \sqrt{ \frac{1}{N} \sum_{m=1}^N \left(\ell^*_{m} - \overline{\ell^*}\right)^2 },
\end{equation}
where $N$ is the number of bootstrap realisations, $\ell^*_m$ the  intensity for a bootstrapped sample, and
$\overline{\ell^*}$ its mean over all realisations. We also found the ``signal-to-noise ratio'' for each grid point:
\begin{equation}
    G = \frac{\ell}{\sigma_{\ell}}.
    \label{eq:sn}
\end{equation}

\section{Description of the catalogue}
\label{app:desc}
The catalogue consists of several tables with some redundancies between them. For each density level $D$ there
exists a table with all superclusters found at that threshold. These tables contain the following information (some less
important properties are omitted here, but can be found in the \verb1readme1 files): 
\begin{itemize}
    \item an unique identification number in the long and short forms;
    \item the number of galaxies and groups (the latter for the main sample alone);
    \item the supercluster volume as the number of the constituent grid cells times the cell volume (Eq.~\ref{eq:vol});
    \item the supercluster luminosity as the sum of densities at grid vertices (Eq.~\ref{eq:ldf});
    \item the supercluster luminosity as the sum of the observed galaxy luminosities (Eq.~\ref{eq:lum});
    \item the supercluster luminosity as the sum of the weighted galaxy luminosities (Eq.~\ref{eq:wlum}). For the main
sample supercluster catalogue, we consider this as the best estimate of the total luminosity of the supercluster;
    \item the maximum density in the supercluster;
    \item the equatorial coordinates (J2000 here and hereafter) and the comoving distance of the highest density peak;
    \item the equatorial coordinates and the comoving distance of the centre of mass (Eq. \ref{eq:massc});
    \item the cartesian coordinates (Eq.~\ref{eq:xyz}) of the highest peak and of the centre of mass;
    \item the supercluster diameter as the maximum distance between the galaxies in the supercluster;
    \item the identifier of the ``marker'' galaxy in the \citet{tago2010} catalogue;
    \item the equatorial coordinates and the redshift of the ``marker'' galaxy;
    \item the confidence estimate for the supercluster found from the signal-to-noise field $G$ (Eq.~\ref{eq:sn});
    \item shows if a supercluster is in contact with the mask boundary (1 -- yes, 0 -- no);
    \item the number of objects that will split from the supercluster above the current density threshold.
\end{itemize}
A similarly structured supercluster catalogue with adaptively assigned density thresholds has been compiled by combining
the supercluster data in the tables described above. For each supercluster we take the data from the fixed level
catalogue that corresponds to its defining density level and add the threshold value.

Additionally, we provide lists of galaxies and groups, together with the supercluster identifiers they are attributed
to, for all density levels. We also present the supercluster splitting tree in the form of a table, where each
supercluster is given the identifier of the object it belongs to at all given thresholds.

As the full volume of the supercluster catalogues is very large, we have chosen to upload only a part of them to the
CDS. There are the two adaptive catalogues, one for the main sample and the other for the LRGs, and two fixed-level
catalogues, of $D=5.0$ for the main sample and of $D=4.4$ for the LRGs. The full catalogue is accessible at:
\url{http://atmos.physic.ut.ee/~juhan/super/} with a complete description in the \verb1readme1 files.

\end{appendix}

\end{document}